# Mechanical buckling induced periodic kinking/stripe microstructures in mechanically peeled graphite flakes from HOPG

Manrui Ren [1], Ze Liu [1], Quan-shui Zheng [1*], Jefferson Zhe Liu [2*]

[1] Department of Engineering Mechanics and Center for Nano and Micro Mechanics, Tsinghua University, Beijing 100084, China

[2] Department of Mechanical and Aerospace Engineering, Monash University, Clayton, VIC 3800, Australia

*Email: zhe.liu@monash.edu and qszheng@tsinghua.edu.cn

**Abstract**

Mechanical exfoliation is a widely used method to isolate high quality graphene layers from bulk graphite. In our recent experiments, some ordered microstructures, consisting of a periodic alternation of kinks and stripes, were observed in thin graphite flakes that were mechanically peeled from highly oriented pyrolytic graphite (HOPG). A theoretical model is presented in this paper to understand the formation of such ordered microstructures, based on elastic buckling of a graphite flake being subjected to a bending moment. The width of the stripes predicted from this model agrees reasonably well with our experimental measurements.

**Introduction**

Graphene, as a single layer of carbon atoms connected by $sp^2$ bonds in a hexagonal lattice, has attracted extensive research in the past decade, because of its remarkable mechanical, electrical, and thermal properties as well as its excellent biocompatibility. It has the highest Young's elastic modulus [1], superior carrier mobility [2-4], and excellent thermo-conductivity [5]. These excellent properties render graphene as a promising candidate in nano-electro-mechanical and micro-electro-mechanical systems (NEMS/MEMS) such as mechanical oscillators [6-9] and electromechanical actuators [10-17], nanofluidic devices [18-20], circuit elements in electronic devices [21], the transparent electrodes in photovoltaic and LCD applications [22] and so on.

There are several methods to produce the graphene layers, *e.g.*, mechanical exfoliation of highly oriented pyrolytic graphite (HOPG) [23], surface segregation [24], chemical vapor deposition [25], thermal decomposition of SiC [26], chemical reduction of graphite oxide [25,27], and so on. Among these methods, the mechanical exfoliation of HOPG can produce a graphene layer with a much higher crystal quality than other methods. Actually the first graphene was produced using this method [23].

In our previous mechanical exfoliation experiments of HOPG [28], we observed some interesting ordered microstructures that were composed of periodic alternation of stripe and kink in the peeled thin graphite flakes. The details of experiments are reported in Ref. 28. **Figure 1**(a) shows a typical graphite flake produced in our experiments. The optical microscope image of top-view of the flake [**Figure 1(b)**] shows a series of wide stripes separated by some narrow dark folding lines. The width of the stripes is fairly uniform, about 80-100 μm. **Figure 1(c)** shows the SEM image of the side-view of the graphite flake. The folding lines observed in **Figure 1(b)** are actually some kinking structures. The stripe width in different flakes varies from 20 to 120 μm. Apparently there is a correlation between the stripe width and the flake thickness.

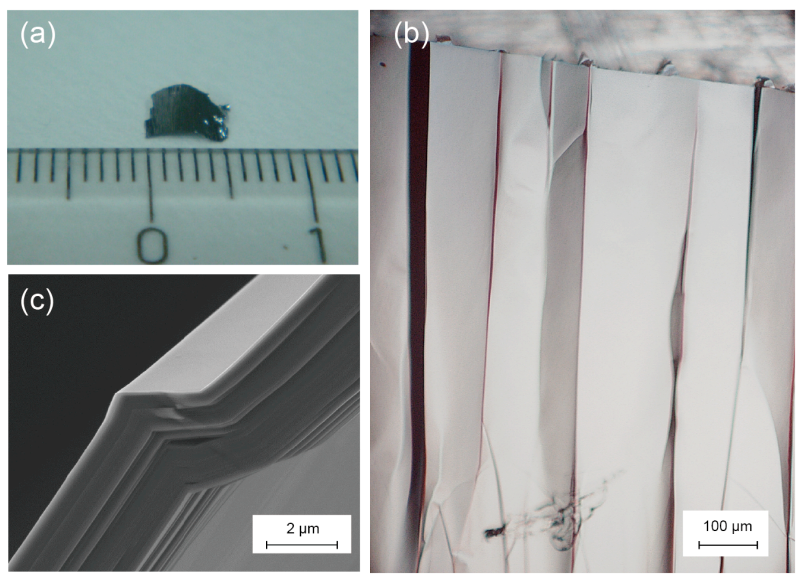

**Figure 1.** (a) A typical graphene flake fabricated using mechanical exfoliation from a HOPG. (b) Top-view of the graphite flake (Optical Microscope, 350 times). (c) SEM image of side-view of the folding lines.

In this paper, the formation of such a periodic microstructure is attributed to the alternation of two mechanical processes during the exfoliation: (1) peeling of a graphite flake and (2) mechanical buckling of the flake being subjected to a bending moment. Theoretical models are presented to describe these two processes and thus to understand the formation of the observed ordered microstructures. The widths of the stripes predicted from our models agree reasonably with the experimental measurements.

**Theoretical model for mechanical peeling**

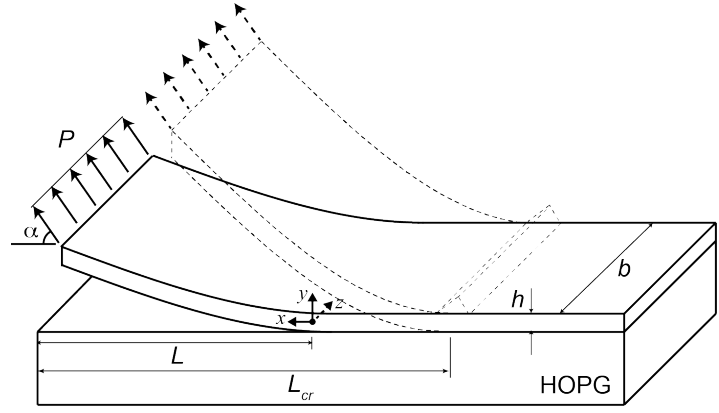

**Figure 2.** Sketch to demonstrate the mechanical exfoliation process. A thin graphite flake is peeling off from the HOPG substrate subject to the peeling force *P* that has an angle *α* from the *x*-axis (*i.e.*, from the HOPG substrate surface). The dashed plot sketches the peeled flake with a buckling/kinking structure forming at the detaching front line, being subject to a mechanical bending. The buckling takes place when the peeled length *L* reaches a critical value $L_{cr}$.

**Figure 2** depicts the peeling process in the mechanical exfoliation of HOPG. A load is applied at one edge of the graphite flake (using a scotch tape) with the other end adhered to the HOPG substrate. We assume that there is an angle *α* between the peeling force *P* and the *x*-axis. The deformation of the half-peeled graphite flake can be described as a cantilever beam/plate being subjected to the peeling load. We model the flake as a Timoshenko beam/plate. The governing equations are

$$\begin{cases} M = P_y(L-x) = EI\dfrac{d\varphi}{dx} \\ \dfrac{dM}{dx} = -P_y \\ P_y = \kappa GA\left(\dfrac{dw}{dx} - \varphi\right) \end{cases}, \quad (1)$$

where *P* is the applied peeling force, *M* is the resultant bending moment, *w* is the displacement of the mid-surface, *E* is Young's modulus, *G* is shear modulus, *I* is the bending moment of inertia, *A* is the cross-section area, *φ* is the angle of rotation of the normal to the mid-surface of the beam, and *κ* is the Timoshenko shear coefficient ($\kappa = 5/6$ for a rectangular cross-section). In this paper, the depth of the plate *b* is taken as a unit length. The boundary conditions are

$$\begin{aligned} w(0) &= 0 \\ \left.\dfrac{dw}{dx}\right|_{x=0} &= 0 \\ \varphi(0) &= 0 \end{aligned} \quad (2)$$

The transverse shear deformation is considered because HOPG has a small shear modulus *G* =

4.5 GPa [29] in comparison with the Young's modulus $E$ = 1000 GPa [29,30] along the longitudinal direction of the beam/plate. Solving Eqn. (1) and (2) yields the vertical displacement at the free end of the beam/plate as

$$\Delta = \frac{P_y L^3}{3EI} + \frac{P_y L}{\kappa GA}. \tag{3}$$

Strain energy of the flake can be expressed as $U=P_y\Delta/2$. The external force potential energy is $W = P_y\Delta$. The mechanical potential energy of the system is thus $\Pi = U-W = -U$. Note that the potential energy arising from the stretching force $P_x$ is neglected: $-P_x^2 L/2AE$, because it is much smaller than the potential energy $\Pi$ arising from the bending force $P_y$ (owing to the very large Young's modulus $E$ in the longitudinal direction and the slenderness of the flake $L/h \gg 1$). The creation of new surface due to increase of the length $L$ leads to increase of surface energy. By assuming the peeling process is quasi-static, the virtual decrease of the mechanical potential energy should be equal to the virtual increase of the surface energy in terms of the change of $\delta L$, i.e.,

$$-\delta \Pi = \delta U = 2\gamma \delta L. \tag{4}$$

It then yields the peeling force $P$ as a function of length $L$ as

$$P_y^2 = 4\gamma \left( \frac{L^2}{EI} + \frac{1}{\kappa GA} \right)^{-1}. \tag{5}$$

The maximum bending moment in the cantilever beam/plate occurs at the fixed end,

$$M_{\max} = P_y L = 2\gamma^{1/2} \left( \frac{1}{EI} + \frac{1}{\kappa GAL^2} \right)^{-1/2}. \tag{6}$$

In our experiment, the typical thickness of graphite flake $h$ is measured to be about 1 to 2 μm. The surface energy of graphite $\gamma$ is a not well-determined physical quantity. There are no direct experimental measurements. Some theoretical models were presented to estimate the binding energy (two times of surface energy $\gamma$) of graphite based on some indirect experiments. The value of binding energy is estimated as 51 meV/atom using experimental measured exfoliation energy of a single graphene layer from graphite [31]. Estimate based on measurements of collapsed nanotubes leads to 33 meV/atom [32]. From desorption experiments of aromatic molecules from a graphite surface, Zacharia estimates the binding energy as 62 meV/atom [33]. Recent quantum Monte Carlo computations report the binding energy as 56 meV/atom [34]. Our recent experiments that measured the deformation of mesoscopic graphite flakes spanning on graphite steps lead to an estimate about 31 meV/atom [35,36]. If we take surface energy $\gamma$ = 0.168 J/m$^2$ (corresponding to binding energy 55 meV/atom), $h$ = 1 or 2 μm, and $\alpha$ = 90°, **Figure 3** shows the peeling force $P$ and the maximum bending moment $M_{\max}$ in the peeled flake as functions of peeled length $L$.

In **Figure 3**, the obtained peeling forces decrease with the increase of the peeled length. This is easy to understand by analogous to the case of interface crack propagation. In contrast to the peeling force, the maximum bending moments in the graphite flakes increases with the peeling length $L$, approaching to $2\sqrt{EI\gamma}$ asymptotically [Eq. (6)]. Also a thicker graphite flake requires a larger peeling force and bending moment. This is due to the higher bending stiffness $EI$ and cross-section area $A$ [Eq. (5) and (6)].

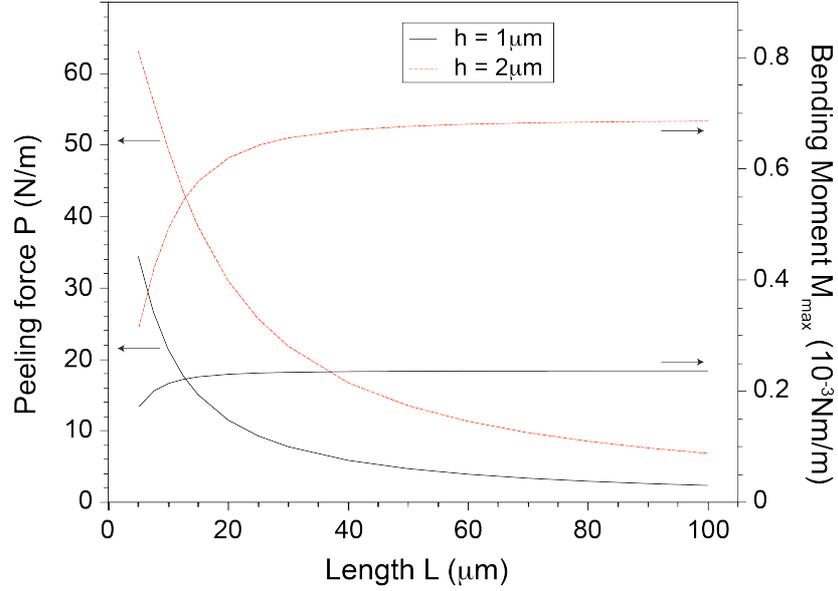

**Figure 3.** The peeling force $P$ and the maximum bending moment $M_{max}$ in graphite flakes versus the peeled length $L$ (Eq. (5) and (6)) at thickness $h$ = 1 or 2 $\mu$m.

**Mechanical buckling of graphite flakes**

The unique multilayer structure of a graphite or a miultiwalled carbon nanotube (MWCNT), *i.e.*, high in-layer modulus/stiffness and extremely low interlayer shear stiffness, gives rise to the easy sliding between adjacent layers/tubes when subjected to compression load, resulting in a mechanical buckling [37-40]. In experiments, it has been observed that the rippling structures appear in the compressive side of a MWCNT under bending [1]. In **Figure 3**, the maximum bending moment (at the peeling front line) increases with the peeled length $L$. It can be expected that once the compressive strain reaches a critical value, $\varepsilon_{cr}$, the buckling will occurs at this point and thus results in the kinking microstructure (**Figure 1**). It is well-known that mechanical buckling leads to a loss of capability of a structure to support external loads. So we can assume release of the peeling forces upon the appearance of the mechanical buckling. The peeling process then restarts, driven by the continuingly exerted peeling force $P$. Alternation of the peeling and buckling processes should produce the observed periodic microstructure. The critical length $L_{cr}$, at which the mechanical buckling occurs, is thus the width of the stripes in our graphite flake samples.

For a beam/plate subject to a peeling force $P$ as shown in Figure 2, the maximum compressive strain can be expressed as $\varepsilon_{max} = M_{max}h/2EI - P_x/EA$ where $h$ is the thickness of the flake. At the critical buckling point, $\varepsilon_{max} = \varepsilon_{cr}$, it yields

$$\left(\varepsilon_{cr}^2 - \frac{12\gamma}{Eh}\right)L_{cr}^2 + \frac{4\gamma \cot\alpha}{E}L_{cr} + \frac{Eh^2\varepsilon_{cr}^2}{12\kappa G} - \frac{\gamma h \cot^2\alpha}{3E} = 0 \quad (7)$$

The critical length $L_{cr}$ is thus:

$$L_{cr} = h\left[\frac{2\gamma \cot\alpha}{12\gamma - Eh\varepsilon_{cr}^2} + \frac{Eh\varepsilon_{cr}^2}{12\gamma - Eh\varepsilon_{cr}^2}\sqrt{\frac{E}{12\kappa G}\frac{12\gamma - Eh\varepsilon_{cr}^2}{Eh\varepsilon_{cr}^2} + \frac{\gamma \cot^2\alpha}{3Eh\varepsilon_{cr}^2}}\right] \quad (8)$$

If the force $P$ is applied vertical to the substrate, i.e., $\alpha = 90^\circ$, we have

$$L_{cr} = h\sqrt{\frac{E}{12\kappa G}}\left(\frac{Eh\varepsilon_{cr}^2}{12\gamma - Eh\varepsilon_{cr}^2}\right)^{1/2}. \quad (9)$$

Our model shows that the $L_{cr}$ depends on the thickness of graphite flakes $h$ and other physical properties of HOPG materials.

The critical mechanical buckling strain $\varepsilon_{cr}$ of ideal graphite or multi-walled carbon nanotube structures were studied previously using different continuous mechanical models [37-39]. The obtained value varies from 0.3% to 0.6%. Considering that the defect could significantly reduce the critical buckling strain, the $\varepsilon_{cr}$ value in experiments could be much smaller. The surface energy determined from different theoretical models and indirect experiments also have a large variation as discussed earlier. Thus, **Figure 4** and **5** depict the $L_{cr}$ as a function of the thickness of graphite flakes $h$ for a set of different $\alpha$, $\varepsilon_{cr}$, and $\gamma$. Three different angle values of $60^\circ$, $90^\circ$, and $120^\circ$ are examined, which seem reasonable considering our experimental setups [28]. We selected the critical strain $\varepsilon_{cr}$ between 0.05% and 0.3%. As for the surface energy $\gamma$, we selected 0.095 J/m$^2$ < $\gamma$ < 0.189 J/m$^2$, which is consistent with the range as determined in other studies. Our theoretical results are also compared with our experimental results in **Figure 4** and **5**. The experimental results were measured from the SEM images of the side-views of graphite flakes.

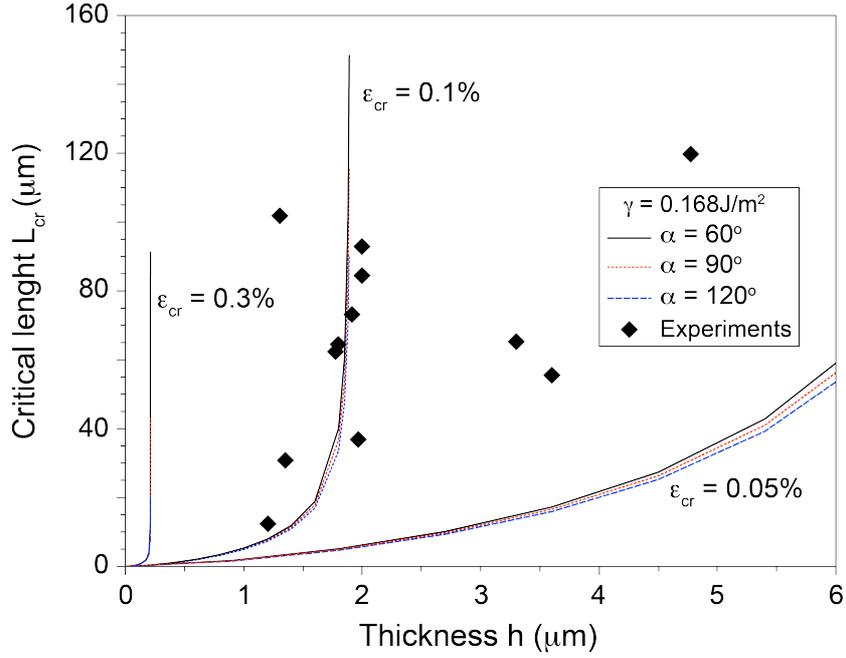

**Figure 4.** The critical buckling length $L_{cr}$ as a function of graphite flake thickness $h$ at angle $\alpha$ of 60°, 90°, or 120° and critical buckling strain $\varepsilon_{cr}$ = 0.05%, 0.1% or 0.3%. The surface energy is taken as 0.168J/m².

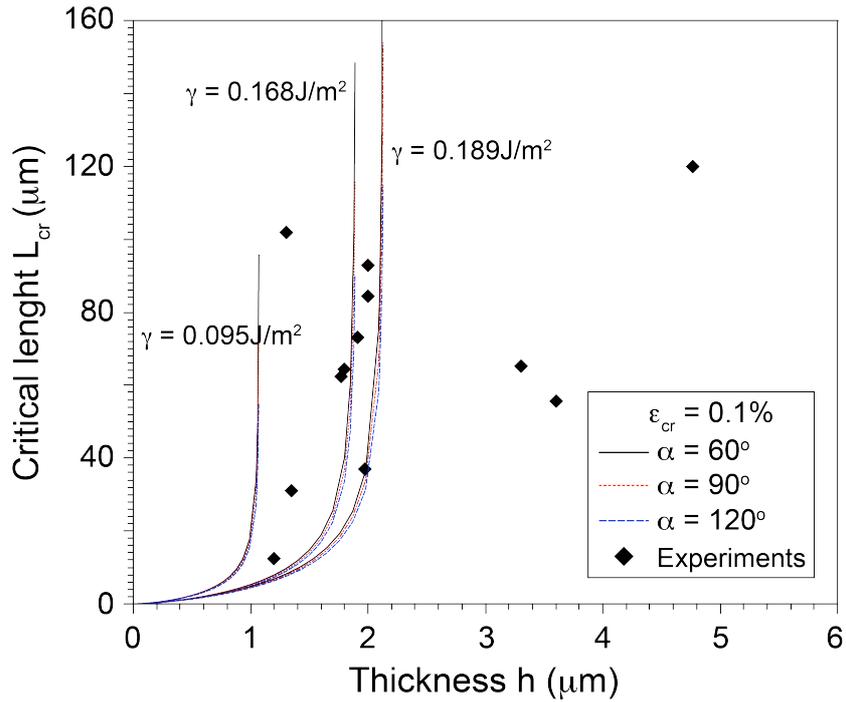

**Figure 5.** The critical buckling length $L_{cr}$ as a function of graphite flake thickness $h$ at angle $\alpha$ of 60°, 90°, or 120° and surface energy $\gamma$ = 0.095, 0.168, 0.189J/m². The critical buckling strain is taken as 0.1%.

In **Figure 4** and **5**, given the set of parameters $\alpha$, $\varepsilon_{cr}$, and $\gamma$, the $L_{cr}$ increases with $h$, which agrees quite well with the trend of our experimental results. From Eq. (6), at a given peeled length $L$, the maximum bending moment in the peeled flake is scaled as $\sim h^{1.5}$. The maximum compressive strain in the flake should scale as $\varepsilon_{max} \sim M_{max}h/EI \sim h^{-0.5}$. For a thicker flake, owing to the increase of bending stiffness $EI$, the $\varepsilon_{max}$ is smaller than that of a thinner flake. Thus the buckling of a thicker flake would take place (*i.e.*, $\varepsilon_{max} \geq \varepsilon_{cr}$) at a longer peeled length $L$ (because $M_{max}$ increases with $L$ as shown in **Figure 3**). It should be noted that in **Figure 4** and **5**, the $L_{cr}$ approaches to infinity at some critical $h$ values. It suggests that for graphite flakes with a thickness higher than these critical $h$ values, buckling will never happen. Mathematically we found that imaginary results of $L_{cr}$ were obtained from Eq. (8) and (9) for this case. In our experiments, within the region $1\mu m < h < 2\mu m$, the measured $L_{cr}$ results exhibit a significant increase with the increase of $h$, which appears to support the theoretical model.

**Figure 4** shows a sensitive dependence of $L_{cr}$ on the selection of $\varepsilon_{cr}$. Overall a smaller $\varepsilon_{cr}$ value leads to a smaller $L_{cr}$, which is easy to understand. Taking $\varepsilon_{cr} = 0.3\%$ yields a significant overestimation of $L_{cr}$. Actually it predicts that graphite flakes with a thickness $h > 0.22\mu m$ will not buckle, which clearly is inconsistent with experiments. The $L_{cr}$ results predicted using a smaller $\varepsilon_{cr}$ value, *i.e.*, $0.1\%$, appears to agree with most of the experimental results quite well, particularly near $h \sim 1–2\mu m$. We believe $\varepsilon_{cr} \sim 0.1\%$ is a reasonable estimate for the buckling critical strain in experiments. Note that some of our experimental results near $3\ \mu m < h < 5\ \mu m$ are much lower than the theoretical model predictions using $\varepsilon_{cr} = 0.1\%$. Reducing $\varepsilon_{cr}$ to $0.05\%$ corrects the trend of disagreement. It is possible that these graphite flakes were more defective and thus a smaller $\varepsilon_{cr}$ value should be adopted.

**Figure 5** shows a moderate effect of interface energy $\gamma$ on determining the critical length $L_{cr}$ in our model. For the values examined (corresponding to previously reported data), the predicted range of $L_{cr}$ covers most of our experimental data. To peel a strongly bonded interface (a higher $\gamma$), we need a larger external force $P$ [Eq. (5)], yielding a larger bending moment $M$ [Eq. (6)] and then a higher $\varepsilon_{max}$. Therefore the graphite flakes prone to buckle at a shorter $L$ as shown in **Figure 5**.

Utilizing mechanical buckling can alter the surface morphology of thin films and thus modulate the surface physicochemical properties, giving rise to various applications, such as dynamically controlled surface wettability [41], enhancement of light extracting efficiency from organic light-emitting diodes [42], and micro-fluidic devices [43] and artificial muscle actuators [41]. The observed periodic microstructures in our mechanically peeled graphite flakes could find promising applications in these fields. Our mechanical model presented in this paper may provide some valuable design guidelines. Our model [Eq. (8)-(9) and Figure (4)-(5)] predicts that reducing thickness $h$ should give rise to a shorter stripe width $L_{cr}$. This conclusion is very appealing since it could provide an effective means to fabricate graphite surfaces with densely distributed microstructures, which is critical for applications such as controllable surface wettability and micro/nano-fluidic devices.

In summary, we present a continuum mechanical model to describe the mechanical exfoliation process of the HOPG and to understand the formation of the periodic kinking/stripe microstructures observed in our experiments. The mechanical exfoliation of HOPG includes the repetition of two alternation processes: mechanical peeling and mechanical buckling of the graphite flakes subject to bending. The predicted stripe width agrees with the experimental measurements reasonably well. It is observed that the surface energy $\gamma$ has a moderate effect in our model to determine the strip width. We find that the predicted strip width sensitively depends on the critical buckling strain $\varepsilon_{cr}$ value. A comparison with experimental results suggest $\varepsilon_{cr} \sim 0.1\%$. This value is smaller than those reported previously for ideal graphite structures or multi-walled carbon nanotubes. This difference could be attributed to the defects in the HOPG samples in experiments.

## Acknowledgements

Q.S.Z. acknowledges the financial support from NSFC through Grant No. 10832005, the National Basic Research Program of China Grant No. 2007CB936803, and the National 863 Project Grant No. 2008AA03Z302. J.Z.L. acknowledges the support from engineering faculty of Monash University through seed grant 2014.